\documentclass[onecolumn]{svjour3}
\usepackage{graphicx}
\usepackage{latexsym}
\usepackage{color}
\usepackage{makeidx}
\usepackage[english]{babel}
\usepackage{braket}
\usepackage{amsmath}
\usepackage{cite}

\begin{document}

\authorrunning{J.Klamut and T. Gubiec}
\titlerunning{Directed Continuous-Time Random Walk with memory}
\title{Directed Continuous-Time Random Walk with memory}
\author{Jaros{\l}aw Klamut \and Tomasz Gubiec } 
\institute{Tomasz Gubiec
\at Center for Polymer Studies and Department of Physics, Boston University, Boston, MA 02215, USA
\and
Tomasz Gubiec \and Jaroslaw Klamut
\at Faculty of Physics, University of Warsaw, Pasteur Str. 5, 02093 Warsaw, Poland}
\date{Received: date / Revised version: date}

\maketitle
\begin{abstract}
We propose a new Directed Continuous-Time Random Walk (CTRW) model with memory. As CTRW trajectory consists of spatial jumps preceded by waiting times, in Directed CTRW, we consider the case with only positive spatial jumps. Moreover, we consider the memory in the model as each spatial jump depends on the previous one. Our model is motivated by the financial application of the CTRW presented in [Phys. Rev. E 82:046119][Eur. Phys. J. B 90:50]. As CTRW can successfully describe the short term negative autocorrelation of returns in high-frequency financial data (caused by the bid-ask bounce phenomena), we asked ourselves to what extent the observed long-term autocorrelation of absolute values of returns can be explained by the same phenomena. It turned out that the bid-ask bounce can be responsible only for the small fraction of the memory observed in the high-frequency financial data.
\end{abstract}

\section{Introduction} \label{sec_intro}
In 1956 two physicists Montroll and Weiss, in the context of dispersive transport diffusion, introduced a new stochastic process they named Continuous-Time Random Walk (CTRW) \cite{montroll1965}. As dynamics of many complex systems can be described by discrete spatiotemporal events, i.e., the spatial jump of stochastic process preceded by waiting time, the formalism of CTRW seems a natural description. On the other hand, CTRW can be considered as a way to introduce finite, continuous and fluctuating interevent times into a random walk.\\
\newline
Since its introduction, the elegant and flexible concept of CTRW found many applications and inspired at least three generations of scientists. It is worth to mention that recently The European Physical Journal B published a special issue titled ''Continuous Time Random Walk fifty years on''. The extended introduction to this topical issue by Kutner and Masoliver lists all applications and extensions of CTRW created up to 2017 \cite{Kutner2017}.\\
\newline
CTRW was initially introduced to describe a photocurrent relaxation in amorphous films \cite{SM, pfister1978, shlesinger1984, weiss1994, bouchaud1990}. A broad spectrum of other applications and arrangements contains: diffusion in probabilistic fractal structures (percolations clusters \cite{ben2000} and fractal diffusion \cite{HILFER200335}), aging of glasses \cite{EB,MB}, nearly constant dielectric loss in disordered ionic conductors \cite{dieterich2009}, cardiological rhythms \cite{iyengar1996}, electron transfer \cite{nelson1999}, search models \cite{lomholt2008}, transport in porous media \cite{margolin2000}, diffusion of epicenters of earthquakes aftershocks \cite{HS}, subsurface tracer diffusion \cite{scher2002}, hydrogen diffusion in nanostructure compounds \cite{hempelmann1999} or even human travel \cite{hufnagel2006}. In this work we are particularly interested in CTRW models used in the description of financial markets, mainly financial time series, where the dependencies and distributions of times between transactions and price changes are considered \cite{f1, f2, f3, f4, f5, f6, f7, f8, f9, f10, kutner2002, scalas2006, perello2008, Gubiec2017, TG_1, kasprzak2010}.\\
\newline
In the majority of cases, the analyzed CTRW models focus on the spatial distribution with zero mean or even symmetric distribution. In other words, the drift term is usually neglected. The case of drift was studied in \cite{metzler2000} (and references therein). The case of canonical CTRW, where both spatial and temporal distributions are i.i.d. and they do not depend on each other, turns out to be a compelling model, able to describe many cases of normal or anomalous diffusion. Different types of CTRW are obtained if mean waiting time is finite or diverging (but assuming finite variance of the spatial distribution). In the first case, we observe a normal diffusion, in the latter subdiffusion occurs \cite{scher1991}. If the variance of the spatial distribution diverges and waiting time distribution has a finite mean, we obtain the description of L\'evy flights.\\
\newline
The other promising branch of CTRW models is the one considering memory, i.e., the dependence between successive jumps. Different types of dependencies were already studied: the backward or forward correlations between spatial jump directions \cite{haus1987} in the case of concentrated lattice gas for the study of the tracer coefficient \cite{kehr1981}, even taking into account the dependencies over several subsequent jumps \cite{kutner1985}. Also, models driven by the negative feedback in consecutive jumps were built, considering one-step memory \cite{TG_0, TG_1} and later two-step or even infinity-step memory \cite{Gubiec2017}. Their potential applications cover the Le Chatelier-Braun principle of contrariness. Memory in waiting times also appeared in some CTRW models \cite{montero2007, montero2011, metzler2010, zebrowski2012, sokolov2009, metzler2013}. Examples of used dependencies are correlations which solely depends on the sign of consecutive jumps \cite{montero2007}, random walk of waiting times \cite{zebrowski2012, metzler2010}, exponential and slowly decaying persistent power-law correlations\cite{sokolov2009}.\\
\newline
Our work is directly motivated by the application of CTRW in the description of high-frequency financial data. The universal properties of all financial price time series are sometimes referred to as stylized facts \cite{Tsay, RCont2001}. There are two well known stylized facts about autocorrelation of price time series. The first one states that the time-dependent autocorrelation of price increments (or logarithmic returns) is negative and quickly decays to zero \cite{montero2007}. A CTRW model with memory \cite{TG_1} successfully reproduced this fact. The second stylized fact states that autocorrelation of the absolute value of price increments (or absolute values of log returns) is a positive slowly decaying function. Also, the amplitude in the second case is usually an order of magnitude higher than in the first case. It is the reminiscence of the so-called volatility clustering phenomenon \cite{RCont2005}. It seems natural to ask if the CTRW model with memory introduced in \cite{TG_1} adapted to describe absolute values of price changes can successfully reproduce the second mentioned stylized fact. We are answering this question below.\\
\newline
The paper is organized as follows: in Section \ref{sec_model} we present the motivation of our work and define and solve the proper stochastic process. In Section \ref{sec_vaf} we obtain Velocity Autocorrelation Function (VAF) and in Section \ref{sec_emp} the comparison with empirical data is made. The intraday-seasonality is taken into account in Section \ref{sec_nonsta}. Finally in Section \ref{sec_sim} we conclude with some additional remarks the results presented in this paper.

\section{Model} \label{sec_model}
We construct a directed continuous-time random walk (CTRW) process with assumptions analogical to ones used in \cite{TG_1} but focused on the absolute values of spatial jumps. This process models stock prices, the value of the process at time $t$ represents the stock price at the corresponding time. A change of its value, called jump, is the price change (which happens immediately when transaction occurs). Waiting time can be interpreted as the time between transactions. We consider one-step memory for consecutive jumps and no dependence between waiting times or between waiting times and jumps. To consider modules of jumps, we create a new process based on the original one. We insert jump length modules in the place of jumps. We obtain directed process, where one-step memory for consecutive jumps is given by
\begin{equation}
\label{h_one_step}
H(R_n|R_{n-1})=(1-\epsilon)H(R_n)+\epsilon\delta(R_n-R_{n-1}),
\end{equation}
where $H(R_n)$ and $H(R_n|R_{n-1})$ are respectively distribution of jump modules and conditional distribution of jump modules. Parameter $\epsilon$ describes the strength of the memory, for $\epsilon=0$ we obtain the model without memory. Considering directed CTRW of absolute values of price changes, Dirac delta describes the same consecutive jumps, not the opposite ones, as it was the case in \cite{TG_1}. To sum up, our model can be described by the probability density functions of $n$th jump $R_n$ after waiting time $t_n$ conditioned on all previous $R_i$ and $t_i$:
\begin{equation}
\label{rho_one_step}
\rho (R_n,t_n|R_{n-1},t_{n-1}; \dots ;R_1,t_1) = H(R_n|R_{n-1}) \psi(t),
\end{equation}
where $\psi(t)$ represents the waiting time distribution (WTD). Results will be presented for any WTD and in two specific cases.
\\
\newline
We cannot use the same waiting time distribution for the first jump as for other jumps \cite{JT, JKT}. This is because the previous (preinitial) jump might have occured at any time before t = 0. Therefore, we should define 
\begin{equation}
\label{psi_1}
\psi_1(t)=\frac{\int_0^\infty dt' \psi(t+t')}{\int_0^\infty dt'' \int_0^\infty dt' \psi(t'+t'')},
\end{equation}
as the waiting time distribution before the first jump. Moreover, for simplicity of notation it is useful to introduce sojourn probability $\Psi(t) = \int_t^\infty \psi(t') dt'$. Above probabilities can be easily expressed in the Laplace domain:
\begin{equation}
\label{Psi}
\begin{split}
\tilde{\psi}(s)& = \mathcal{L}[\psi(t)],\\
\tilde{\Psi}(s) & = \frac{1-\tilde{\psi}(s)}{s},\\
\tilde{\psi}_1(s) & = \frac{1-\tilde{\psi}(s)}{\braket{t}s}, \\
\tilde{\Psi}_1(s) & = \frac{1-\tilde{\psi}_1(s)}{s},
\end{split}
\end{equation}
where $\mathcal{L}[\cdot]$ denotes Laplace transform and $\braket{t} = \int_0^\infty t \psi(t) dt < \infty$ is expected (mean) waiting time. The intermediate dynamic quantity describing the stochastic process is the stochastic, sharp, $n$-step propagator $Q_n(X,R_n;t|\xi), \; n = 1, 2, \ldots$ This propagator is defined as the conditional probability density that the price, which was initially (at $t = 0$) in the origin value ($X=0$) reached by preinitial jump $\xi$, makes its $n$th jump by $R_n$ from $X-R_n$ to $X$ exactly at time $t$. $\tilde{Q}_n(K,R_n;s|\xi)$ is sharp propagator in the Fourier-Laplace domain. The recursion relation between two successive sharp stochastic propagators can be written for any form of $H(R_n|R_{n-1})$ and $n>1$, as follows:
\begin{equation}
\label{Q_1}
\tilde{Q}_n(K,R_n;s|\xi) = \tilde{\psi}(s) e^{iKR_n} \int \limits_{-\infty}^\infty dR_{n-1} H(R_n|R_{n-1}) \tilde{Q}_{n-1}(K,R_{n-1};s|\xi).
\end{equation}
The first sharp propagator $Q_1(X,R_1;t|\xi)$ can be calculated directly from definition
\begin{equation}
\label{Q_2}
Q_1(X,R_1;t|\xi)=Q_1(X;t|\xi) \delta(X-R_1) = \psi_1(t) H(X|\xi) \delta(X-R_1).
\end{equation}
The following sharp propagators can be calculated using Eq. (\ref{Q_1}). After integrating over $R_n$ we obtain the recursion relation
\begin{equation}
\label{Q_3}
\frac{\tilde{Q}_n(K;s|\xi)}{\tilde{\psi}(s)}=(1-\epsilon)\tilde{H}(K)\tilde{Q}_{n-1}(K;s|\xi) + \epsilon \int \limits_{0}^{\infty}dR_{n-1} e^{iKR_{n-1}}\tilde{Q}_{n-1}(K,R_{n-1};s|\xi).
\end{equation}
Finally, we can write the relation between the soft propagator $P(x,t)$, defined as the probability density that process will be in $x$ at the time $t$ starting from $x=0$ at the time $t=0$, and the sharp propagator $Q(x,t)$ (in the Fourier-Laplace domain)
\begin{align}
\label{P_1}
\tilde{P}(K;s) &= \tilde{\Psi}_1(s) + \tilde{\Psi}(s) \tilde{Q}(K;s),\\
\label{Q_4}
\tilde{Q}(K;s) &= \sum \limits_{n=1}^\infty \tilde{Q}_n(K;s).
\end{align}
To obtain explicit formula for the right hand side of Eq. (\ref{Q_4}), in case of one-step memoru defined by Eq. (\ref{h_one_step}), we use the Z-transform in variable $n$ and recurrence relation (\ref{Q_3}).  The result can be simplit substituted into Eq. (\ref{P_1}) and hence we obtain the soft propagator in the following form
\begin{equation}
\label{P_general}
\tilde{P}(K;s) = \frac{1}{s}-\frac{1-\tilde{\psi}(s)}{\braket{t}s^2}+\frac{[1-\tilde{\psi}(s)]^2}{\braket{t}s^2}\frac{S(K; s)}{1-(1-\epsilon)\tilde{\psi}(s)S(K; s)}.
\end{equation}
where
\begin{equation}
\label{S}
S(K; s) = \sum_{n=1}^\infty (\tilde{\psi}(s) \epsilon)^{n-1} \tilde{H}(nK).
\end{equation}
As a result, the soft propagator in the Fourier-Laplace domain takes a reasonably simple form, however it still contains the function $S$ which is given as an infinite sum. Fortunately, to compute moments of process and autocorrelation of velocity we need to know the corresponding derivatives of the soft propagator at point $K=0$, which can be determined explicitly.\\
\newline
The first and the second moment of the process in the Fourier-Laplace domain are
\begin{align}
\label{moment_1}
\tilde{m}_1(s) &= - i \frac{\partial \tilde{P}(K;s)}{\partial K} \biggr\rvert_{K=0}= \frac{M_1}{\braket{t}s^2}, \\
\nonumber
\tilde{m}_2(s) &=- \frac{\partial^2 \tilde{P}(K;s)}{\partial K^2} \biggr\rvert_{K=0}=\frac{M_2+(1-\epsilon)(2M_1^2-M_2) \tilde{\psi}(s) -\epsilon M_2 \tilde{\psi}^2(s)}{\braket{t}s^2(1-\tilde{\psi}(s))(1-\epsilon \tilde{\psi}(s))} = \\
\label{moment_2}
&= \frac{M_2 (1+\epsilon \tilde{\psi}(s))}{\braket{t}s^2(1-\epsilon \tilde{\psi}(s))} + \frac{2(1-\epsilon) \tilde{\psi}(s) M_1^2}{\braket{t}s^2(1-\tilde{\psi}(s))(1-\epsilon \tilde{\psi}(s))},
\end{align}
where $M_i$ is the $i$-th moment of jump modules distribution $H(R)$. The first moment of the directed process in the time space rises linearly in time $m_1(t) = \frac{M_1}{\braket{t}}t$, exactly like for the process without memory. It is worth to notice that it does not depend on $\epsilon$.

\section{Velocity Autocorrelation Function} \label{sec_vaf}
In the general case, the velocity autocorrelation function (VAF) in the time domain is given by
\begin{equation}
\label{VAF_0}
C(t) = \frac{1}{2} \ddot{m}_2(t) - \dot{m}_1^2(t).
\end{equation}
In the case of the directed CTRW process considered in this manuscript, it takes the form
\begin{align}
\label{VAF_general}
C(t) &= \left( \frac{M_2-M_1^2}{2 \braket{t}} \right) \mathcal{L}^{-1} \left[ \frac{1 + \epsilon \tilde{\psi}(s)}{1-\epsilon \tilde{\psi}(s)} \right] + \frac{M_1^2}{2 \braket{t}} \mathcal{L}^{-1} \left[ \frac{1 + \tilde{\psi}(s)}{1- \tilde{\psi}(s)} - \frac{2}{\braket{t}s} \right],
\end{align}
where $\mathcal{L}^{-1}[\cdot]$ is the inverse Laplace transform. 
To investigate the behavior of VAF in the limits $t \to 0$ and $t \to \infty$, we have to check the behavior in limits $s \to \infty$ and $s \to 0$ of the expressions inside inverse Laplace transforms. 
It is known that for $s \to \infty$,  $\tilde{\psi}(s)$ goes to $0$, while for  $s \to 0$ the approximation $\tilde{\psi}(s) \approx 1 - \braket{t}s$ should be used. 
Therefore, in the limit of long times VAF vanishes and for short times we obtain the variance of the process $C(t) \approx \frac{M_2}{2 \braket{t}} \delta(t)$. 
Normalized VAF $C^n(t)$ has Dirac delta at $t=0$ so $C^n(t) = \frac{2 \braket{t}}{M_2} C(t)$. \\
\newline
To compare our model with empirical data we use two specific WTDs: exponential and double-exponential, with explicit results for both. First one is a simple  distribution and its characteristics match stylized facts of financial time series. Second one can be fitted to empirical data with high accuracy \cite{TG_1} and still allow to obtain analitical VAF from the model. Exponential WTD with the mean waiting time equal to $\braket{t}$ is given as
\begin{equation}
\label{time_one}
\psi(t)=\frac{1}{\braket{t}} \exp{\left(-\frac{t}{\braket{t}} \right)},
\end{equation}
and double-exponential WTD with partial mean waiting times equal to $\tau_1$ and $\tau_2$ and weighting parameter $w$ is given as
\begin{equation}
\label{time_two}
\psi(t)=\frac{w}{\tau_1} \exp{\left(-\frac{t}{\tau_1}\right)} + \frac{1-w}{\tau_2} \exp{\left(-\frac{t}{\tau_2}\right)}.
\end{equation}
Mean waiting time of double-exponential WTD is $\braket{t}=w\tau_1 + (1-w)\tau_2$.\\
\newline
For the exponential WTD, VAF is easily expressed as
\begin{equation}
\label{nVAF_one}
C^n(t)=\delta(t) + \frac{2 \epsilon (1-M)}{\braket{t}} \exp{\left( -\frac{(1-\epsilon)t}{\braket{t}} \right)},
\end{equation}
where
\begin{equation}
\label{moment_M}
M=\frac{M_1^2}{M_2} \in (0;1).
\end{equation}
Although exponential WTD does not describe properly the empirical WTD, one can easily interprete meaning of the parameters. Firstly, for $t>0$ VAF is positive (unlike VAF in \cite{TG_1}) and decreases exponentially. Relaxation time increases and the amplitude reduces with longer mean waiting time. Increasing parameter $\epsilon$ results in higher relaxation time and amplitude, especially for $\epsilon=0$ VAF is non-zero only for $t=0$. It is also noticable, that normalized VAF depends only on a ratio between the first moment squared and the second moment of jumps modules. The amplitude of VAF decreases with increase of $M$.\\
\newline
For the double-exponential WTD, which satisfactorily fits empirical data, normalized VAF is
\begin{align}
\label{nVAF_two}
C^n(t) &= \delta(t) + A_0 e^{-v_0 t} + A_1 e^{-v_1t} + A_2 e^{-v_2t},\\
\nonumber
w_i &= \tau_i^{-1}, \\
\nonumber
v &= ww_1+(1-w)w_2,\\
\nonumber
v_0 &= (1-w)w_1 + w w_2,\\
\nonumber
v_i &= \frac{1}{2} \left[ w_1+w_2-\epsilon v - (-1)^i \sqrt{(w_1+w_2 -\epsilon v)^2-4 w_1 w_2 (1 -\epsilon)} \right], \\
\nonumber
A_0 &= 2\frac{M}{v_0}w(1-w)(w_1-w_2)^2, \\
\nonumber
A_i &= (-1)^i \frac{2 \epsilon (1-M)}{v_1-v_2} [w_1w_2-vv_i],\\
\nonumber
i &\in \{ 1,2 \}.
\end{align}
There are three exponents in this formula, except Dirac delta, all with positive amplitudes. The first one is worth mentioning, it does not depend on $\epsilon$. It implicates, that for directed processes VAF can be non-zero even in the case without memory ($\epsilon=0$). That effect does not occur while considering VAF of the original process. In the limit when two-exponent WTD goes to exponential WTD ($w \to 0$ or $1$) this element vanishes. Two other exponents depend on $\epsilon$ and describe decaying with different rates. Similarly like for exponential WTD, increasing $\epsilon$ results in higher VAF. In the case of spatial changes with zero variance ($M \to 1$), these terms are equal to zero.

\section{Empirical results} \label{sec_emp}
\begin{figure}[ht]
\centering
\includegraphics[width=0.75\textwidth]{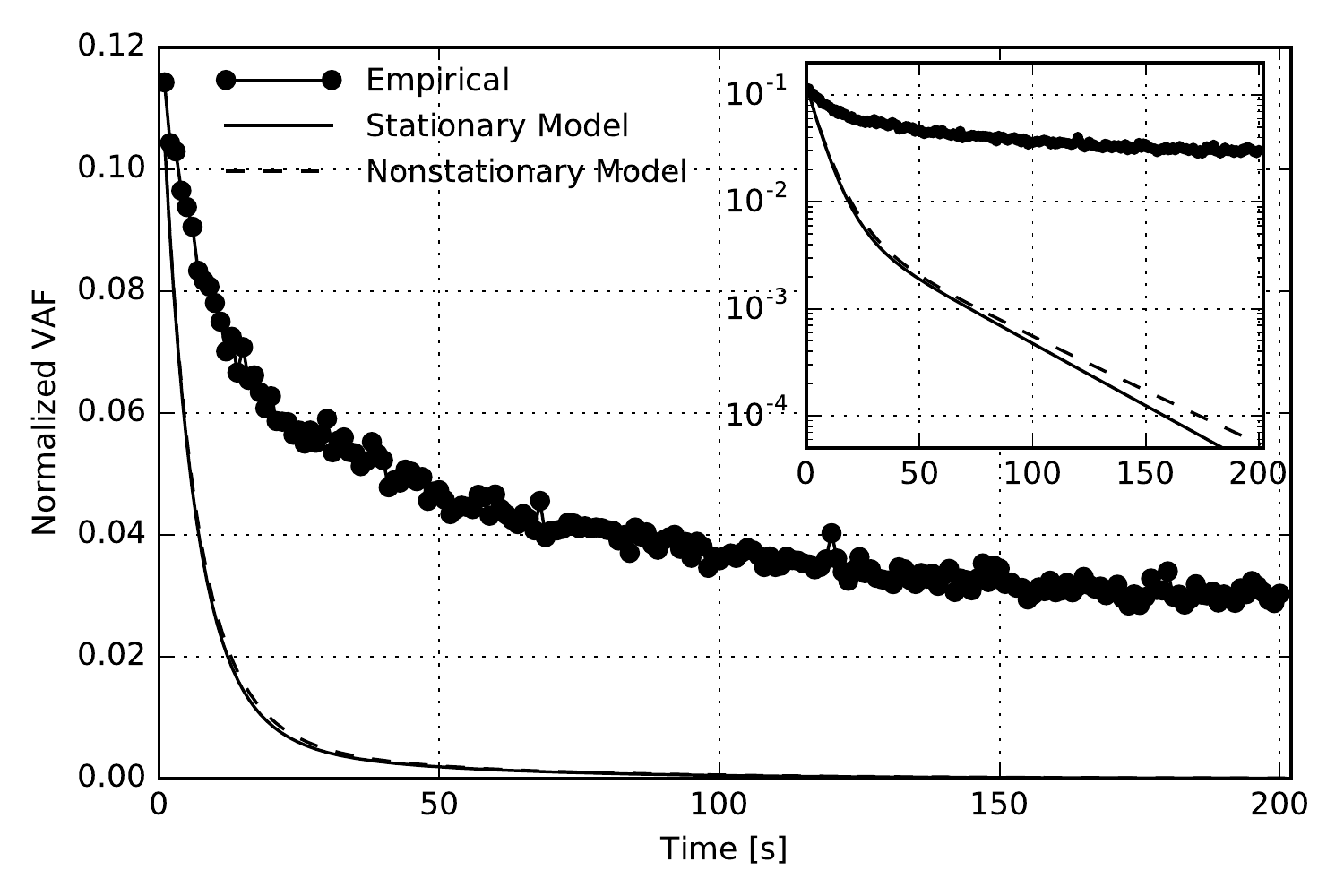}
\caption{Empirical and theoretical VAF of KGHM (one of the most liquid stock at Warsaw Stock Exchange) in years 2011-2012 in the linear and semilogarithmic coordinates. Points are empirical result, solid line represents stationary model and dashed line nonstationary model (see section \ref{sec_nonsta}). Both approaches do not fit empirical data. Fitted values of the parameters are: $M = 0.269,$ $\epsilon = 0.258,$ $\tau_1 = 3.63,$ $\tau_2 = 32.57,$ $w = 0.586,$ $p = 14986,$ $q = 2.25 \cdot 10^8.$}
\label{plot_VAF_1}
\end{figure}

\begin{figure}[ht]
\centering
\includegraphics[width=0.75\textwidth]{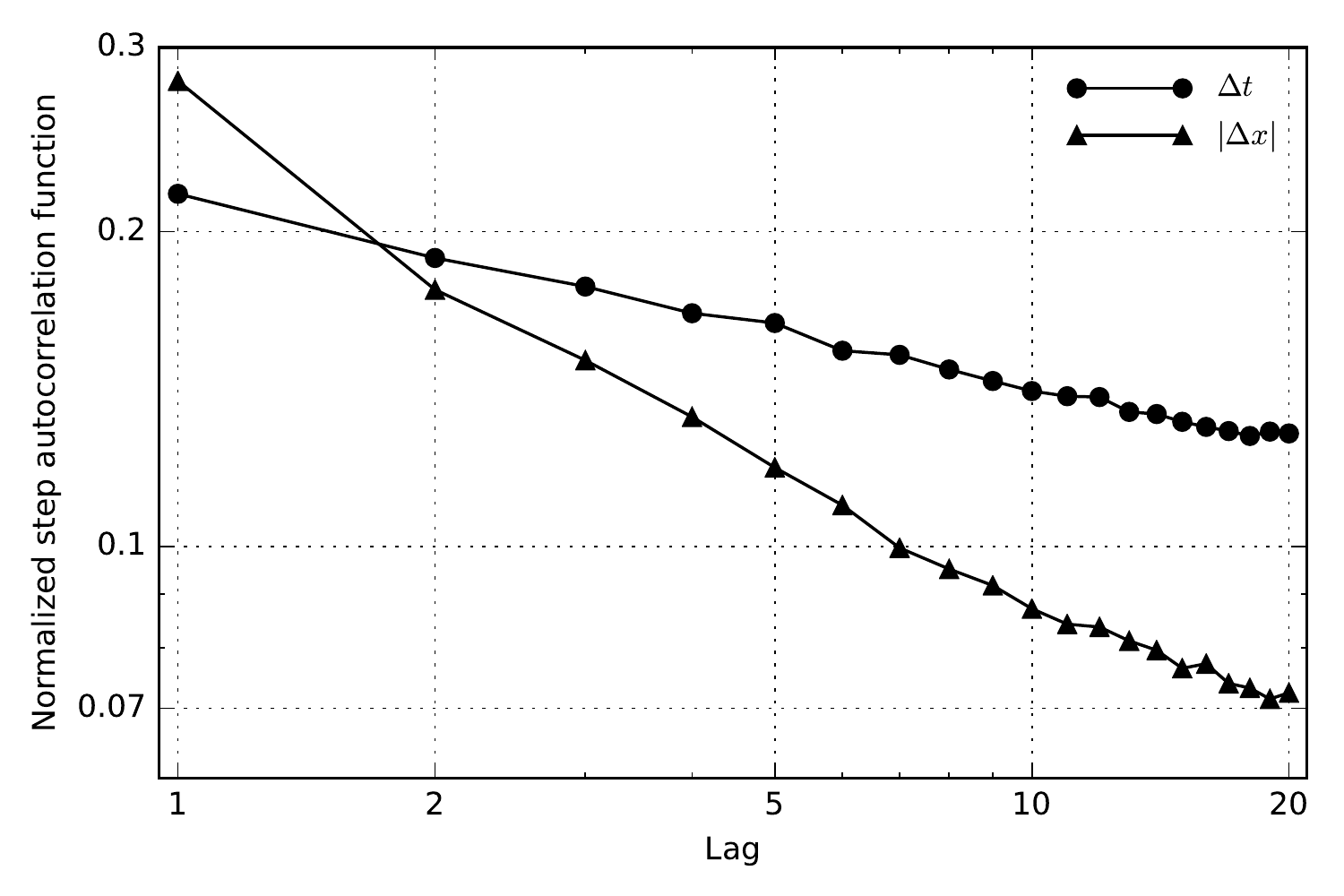}
\caption{Normalized step autocorrelation function of waiting times and the modules of price changes for KGHM in years 2011-2012 in the logarithmic coordinates.}
\label{plot_acf_dt_absdx}
\end{figure}

To compare our model with empirical data, we use tick-by-tick transaction data from Polish stock market (Warsaw Stock Exchange) from years 2011-2012. 
Presented results are calculated for KGHM - one of the most liquid stocks. 
We extract waiting times (periods between transactions) and jumps (price changes) from this data. 
Comparing the model with empirical data requires estimating parameters. 
We obtain $\tau_1$, $\tau_2$, $w$ from fitting two-exponential WTD to empirical histogram using least squares method. 
We calculate two first moments of price changes absolute values $M_1$ and $M_2$ explicitly from the empirical distribution. 
Parameter $\epsilon$ is calculated as one-step autocorrelation of price changes absolute values. 
Empirical VAF was calculated using the method described in \cite{TG_1}.
We want to remind that VAF obtained from the model proposed in \cite{TG_1} (and its modifications in \cite{Gubiec2017, MWilinski}) built using these parameters fits satisfactorily with empirical data. 
As shown in Fig. \ref{plot_VAF_1} theoretical approach does not explain observed VAF for the modules of price changes. 
VAF of the directed process decays much slower than VAF of the original one (see \cite{TG_1}). 
There can be many reasons explaining obtained disagreement: daily seasonality, long-term jump modules dependencies and long-term waiting times dependencies (see Fig. \ref{plot_acf_dt_absdx}). 
In the next section, we check if taking into account of the first one (daily seasonality) can significantly improve the quality of the description of data. Taking into account the two later cases reacquire a new CTRW models with memory and goes beyond the scope of this article.

\section{Nonstationarity} \label{sec_nonsta}
Varying mean inter-trade time during trading session is a stylized fact observed on stock prices on every market \cite{RCont2001, Hasbrouck2007, DacorognaGencayMullerOlsenPictet2001}. 
This intra-day pattern, often called the ''lunch effect'', is characterized by low volatility and long inter-trade times in the middle of the day. 
On the contrary, activity on the stock market is higher at the beginning and the end of a trading session. 
This effect influences the VAF of the process, as described in \cite{MWilinski}, where the general formula for the impact of this phenomenon is given. 
Following \cite{MWilinski} we describe the daily pattern with the rational function
\begin{equation}
\label{nonstat_rational}
\theta(t)=\frac{1}{a[(t-p)^2+q]},
\end{equation}
where $\theta(t)$ can be interpreted as varying mean of the waiting times distribution during a trading session. 
Now, we can obtain an explicit expression for normalized VAF taking the seasonality into account: 
\begin{equation}
\label{nonstat_nVAF}
\begin{split}
C^n(t) &= \sum_{j=0}^2 A_j J v_j^{-\frac{1}{2}} e^{-v_j\tau_{min}} \left[ \text{erf} \left(\sqrt{v_jJ_0}\right) + \text{erf} \left( \sqrt{v_jJ_k} \right) \right],\\
J &= \frac{1}{2(T-t)} \sqrt{\frac{\pi X}{t}},\\
J_0 &= \frac{t}{X}\left( \frac{t}{2}-p \right)^2,\\
J_k &= \frac{t}{X}\left( \frac{t}{2}-(p-T) \right)^2,\\
\tau_{min} &= \dfrac{\frac{ t ^2}{12} + q}{X}  t ,\\
X &= \frac{T^2}{3} - pT + p^2 + q,
\end{split}
\end{equation}
where $T$ is the length of a day, parameters $p$ and $q$ are fitted to data and come from the rational form of day seasonality and erf is the error function. 
As shown in Fig. \ref{plot_VAF_1}, taking nonstationarity into account slightly improves the results. 
However, considering the day seasonality is not enough to explain empirical VAF for jump modules. 

\section{Conclusions} \label{sec_sim}
We proposed and solved Directed CTRW model with one-step memory in jumps and obtained the analytical equation for propagator, first two moments and velocity autocorrelation function (\ref{nVAF_two}). 
Obtained VAF shows interesting properties: it is positive; it decays exponentially but slower than the VAF for the non-directed process; VAF can be nonzero even without any memory ($\epsilon = 0$) for non-exponential WTD. Next, we considered nonstationarity in the form of rational function (\ref{nonstat_rational}) and obtained analytical VAF (\ref{nonstat_nVAF}).\\
\newline
Presented simple model, despite being analytically solvable, turned out to be unable to describe empirical data.
This result suggests that simple bid-ask bounce phenomenon is not sufficient to explain long memory in financial time series and volatility clustering.
We suggest that taking into account at least one of the existing long memories in jumps modules or waiting times (see Fig. \ref{plot_acf_dt_absdx}) would improve the results.

\bibliographystyle{unsrt}
\bibliography{my}{}

\end{document}